# Genome analysis and pleiotropy assessment using causal networks with loss of function mutation and metabolomics

Azam Yazdani[1*], Akram Yazdani[2], Sarah H. Elsea[3], Daniel J. Schaid[4], Michael R. Kosorok[5], Gita Dangol[6], Ahmad Samiei[7,8]

[1] School of Medicine, Boston University, MA 02118, a.mandana.yazdani@gmail.com
[2] Department of Genetics and Genomic Sciences, Icahn School of Medicine at Mount Sinai, New York, 10029, akramyazdani16@gmail.com
[3] Department of Molecular and Human Genetics, Baylor College of Medicine, Houston, TX 77030, Sarah.Elsea@bcm.edu
[4] Division of Biomedical Statistics and Informatics, Mayo Clinic, Rochester, MN, 55905, schaid@mayo.edu
[5] Department of Biostatistics, University of North Carolina at Chapel Hill, NC 27599, kosorok@bios.unc.edu
[6] Health Science Center, The University of Texas MD Anderson Cancer Center, TX 77030, gitags@gmail.com
[7] Hasso Plattner Institute, 14482 Potsdam, Germany, asppagh@gmail.com
[8] Climax Data Pattern, Boston, MA

[*] Corresponding author: Azam Yazdani, Ph.D.
Boston University, Boston, MA
Tel: 832-258-9296
Email: a.mandana.yazdani@gmail.com

## Abstract

Background: Many genome-wide association studies have detected genomic regions associated with traits, yet understanding the functional causes of association often remains elusive. Utilizing systems approaches and focusing on intermediate molecular phenotypes might facilitate biologic understanding.

Results: The availability of exome sequencing of two populations of African-Americans and European-Americans from the Atherosclerosis Risk in Communities study allowed us to investigate the effects of annotated loss-of-function (LoF) mutations on 122 serum metabolites. To assess the findings, we built metabolomic causal networks for each population separately and utilized structural equation modeling. We then validated our findings with a set of independent samples. By use of methods based on concepts of Mendelian randomization of genetic variants, we showed that some of the affected metabolites are risk predictors in the causal pathway of disease. For example, LoF mutations in the gene *KIAA1755* were identified to elevate the levels of eicosapentaenoate (p-value=5E-14), an essential fatty acid clinically identified to increase essential hypertension. We showed that this gene is in the pathway to triglycerides, where both triglycerides and essential hypertension are risk factors of metabolomic disorder and heart attack. We also identified that the gene *CLDN17*, harboring loss-of-function mutations, had pleiotropic actions on metabolites from amino acid and lipid pathways.

Conclusion: Using systems biology approaches for the analysis of metabolomics and genetic data, we integrated several biological processes, which lead to findings that may functionally connect genetic variants with complex diseases.

## Keywords

Loss of Function, genome analysis, underlying metabolomic relationship, causal network in observational study, structural equation modeling, Mendelian randomization principles, instrumental variable, the G-DAG algorithm

## Background

Lack of knowledge of underlying biological processes in genome wide association studies and disease endpoints has led to a focus on intermediate phenotypes, such as metabolites. Metabolomic profiles are integrated readouts of many biological processes and can functionally connect genetic variants to disease risk factors and complex disease endpoints (1)(2)(3). Furthermore, metabolomics can be used to screen for early disease-related changes (4) and assess effects of external stimuli on living cells (5). Considering metabolites as an intermediate molecular mediator between genes and clinical endpoints offers potential to illuminate mechanisms underlying a specific single nucleotide polymorphism (SNP)/gene, as well strengthen the association of a gene with the intermediate trait (6)(7). Thus, metabolite profiles are ideal intermediate biochemical phenotypes for genome wide association studies (GWAS) (6)(8)(9)(10) and whole genome sequencing studies (11)(12).

Examining metabolomic relationships facilitates understanding of functional links between genetic variants and disease endpoints, which goes beyond traditional association analyses that examine one variable at a time. Analyses that consider the joint effects of multiple traits can have greater statistical power than single-trait analyses of genomic association studies (13)(14)(15). In addition, assigning probabilistic models to expression levels of a group of genes (usually within one pathway) enables development of more accurate classifiers and clusters (16)(17)(18)(19). For high-dimensional metabolomic data, systems approaches based on Mendelian randomization principles can provide an even more comprehensive analysis. To advance a deeper understanding of how genes and metabolites might be interconnected, Yazdani et al. (2016) introduced and applied systems biology approaches that can reveal the underlying relationships and provide insights in metabolomics system (20)(21). We capitalize on this prior work and introduce an approach to improve genome analysis and assess pleiotropic gene actions.

We investigated the effects of genetic variants on the human metabolome in the populations of white and non-white from the Atherosclerosis Risk in Communities (ARIC) study (22). In particular, we focus on loss-of-function (LoF) variants that are predicted to result in a non-viable transcript or a greatly truncated protein product (23). To analyze the LoF-metabolite relationships, we utilized two approaches: 1) a single variant test, and 2) a convex-concave rare variant selection (CCRS) approach (11)(24). This latter approach selects genetic variants through a penalization linear model, which assumes sparsity in genomic associations while considering the local linkage disequilibrium. We then applied a systems approach called Genome Directed Acyclic Graph (G-DAG) (25) to model the metabolomic relationships in causal networks. This G-DAG method assured that the necessary assumptions were met in order to then use structural equation modeling to assess genetic findings. This integrative approach facilitates a mechanistic understanding of how genes and metabolites are related by incorporating and modeling relationships of a large number of metabolites and genetic variants, improves genome-metabolite pathway identification, and identifies genes with multiple functions (i.e. pleiotropic actions). Some of the metabolomic causal pathways identified using the G-DAG algorithm matched the available knowledge (20)(21) and some of the novel findings of the G-DAG algorithm are validated clinically (26).

## Results

To investigate the association of LoF genetic variants with metabolites, we used linear regression to first adjust all metabolites for the covariates age, gender, body mass index, phase (two different time points that the metabolites were measured), and ten principal components (to adjust for population stratification). We then identified mutations in the coding sequence (individually or in aggregate within a gene) with significant effect on metabolites. To achieve this, we used two methods to select strong associations. One approach is a penalized model based on gene-level analyses (simultaneously evaluating all LoF variants in a gene), and is called the CCRS method (Convex-Concave Rare variant Selection). The CCRS method uses a tuning parameter to select LoF-metabolite associations. The set of tuning parameters included 1.5, 0.3 and 0.01, and the tuning parameter that resulted in the minimum BIC was chosen in the final model. The findings are provided in tables S4 and S5 for EA and AA respectively. The second approach was based on single variant tests, and the level of statistical significance for the test was based on Bonferroni correction for 122 metabolites, 451 and 372 LoF mutations in EA (European-American/white) and AA (African-American/non-white) populations respectively. The findings are provided in tables S6 and S7 for EA and AA respectively. We then focused on the variants that were commonly selected by both approaches to reduce the false discovery rate.

The EA and AA participants differed in several ways. The size of covariate effects differed between the EA and AA subjects, the EA and AA individuals were from different geographical regions, and different diets and environment could impact biochemical pathways. Furthermore, ancestry can affect metabolites. In an attempt to reduce confounders, including environmental and regional dietary variations that may impact metabolites, we focused on AA participants from Jackson, Mississippi. We identified the underlying metabolomic relationships separately for the EA and AA groups and we did not expect that the two causal networks EA and AA be identical. However, the two causal networks were very similar with respect to the metabolites with essential roles in the metabolomic system. For example, the relationship between leucine, valine, and isoleucine and the high impact of these three metabolites in the metabolomic system were the same in both causal networks. Among fatty acids, however, the relationships between AA and EA were not very similar. Because differences between EA and AA participants might make it impossible to compare the results between these two populations, we avoided any comparison below.

We integrated the results of LoF-metabolite investigation with the identified metabolomic causal network in each population. To identify underlying relationships among metabolites, we utilized the G-DAG algorithm and identified a metabolomic causal network over 122 metabolites distributed across multiple functional classes in each population. Because the principles of the G-DAG algorithm are based on Mendelian randomization, where genetic variants act as instrumental variables, hence anchoring the direction of causation, the results from this algorithm provide directions of effects in a network. This provided a valid way to subsequently use structural equation modeling, see the method section and (27). Figure 1 illustrates the steps of the analysis. The findings are presented in Tables 1 and 2 for EA and AA, respectively and some of them are validated using an independent sample.

**Genome Analysis of AA Population**

Some of the results from our genome analysis of the AA population are illustrated and described below.

***GPR97***: In the genome analysis, *GPR97* showed significant effects on the metabolites glycocholenate-sulfate, oleate, and eicoseneate from the lipid pathway; therefore, this gene was hypothesized to have pleiotropic action. Looking at the relationships among the metabolites (Figure 2), we observed the metabolites oleate and eicoseneate to have direct relationshisps. Using structural equation models to estimate the effect of *GPR97* on eicoseneat, we found this effect to not be statistically significant. Therefore, we concluded that *GPR97* does not have a pleiotropic action on both metabolites oleate and eicoseneate, but rather *GPR97* directly influences only the oleate metabolite.

***BNIPL***: From the genome analysis, we found that *BNIPL* has significant effects on the two metabolites octanoylcarnitine and decanoylcarnitine from the lipid pathway (sub-pathway carnitine metabolism). Therefore, we hypothesized that *BNIPL* has a pleiotropic effect. From the metabolomic causal network, we found that the two metabolites are directly associated (Figure 3). To assess the hypothesized pleiotropic effect of *BNIPL*, we modeled the relationship between the two metabolites using structural equation models. The results from this analysis showed that effect of *BNIPL* on octanoylcarnitine did not remain statistically significant. Therefore, the pleiotropic hypothesis was rejected.

Note that Figures 2 and 3 both are showing the same causal network. In Figure 2, we focused on the metabolites affected by *GPR97* and in Figure 3, we focused on the metabolites affected by *BNIPL*. In each Figure, we highlighted the relationship between the metabolites of interest and brought those metabolites to the border of the causal network to show the relationships.

**The impact of LoF mutations on AA and EA metabolomics**

Further analyses of the metabolomic causal network identified five modules (densely connected metabolites). We noticed that the identified modules each referred to a set of metabolites with similar function (e.g. metabolites in a pathway) that work together to achieve a coordinated biologic outcome. Below we summarize the effect of LoF mutations on the identified modules.

One of the modules includes hippurate, p-cresol sulfate, methylcatecholsulfate, catecholsulfate, and phenylacetylglutamine, with urea and glutarylcarnitine as neighbors. Investigation of these metabolites indicates that they mostly result from gut metabolism of a diet rich in protein and polyphenols. Many of these metabolites represent major gut contributions to uremic solutes (e.g. methylcatechol sulfate) and uremic toxicity (e.g. p-cresol sulfate) (28)(21). In the AA causal network, the LoF-metabolite findings revealed that these metabolites were highly influenced by genes harboring LoF mutations: phenylacetylglutamine is influenced by *DCLK3* (p=4e-09), *ZSWIM1* (p=4e-15), and *TMPRSS3* (p=2e-16); urea is influenced by *LTK* (p=6e-09), *AVEN* (p=2e-08) and *CLSPN* (p=1e-13), and glutarylcarnitine is influenced by *GUCA1C* (p=4e-11) (seeFigure 4). In contrast, these metabolites were not influenced by any LoF loci in the EA causal network.
In the AA causal network, hormone-related metabolites were significantly influenced by LoF mutations (Figure 5). These hormone-related metabolites did not influence other metabolites in the metabolomic causal network. Therefore, the LoF loci associated with hormone-related metabolites did not influence other metabolites in the metabolomic causal network.

In the EA causal network, the hormone-related metabolites glycocholenate sulfate, pregn steroid monosulfate, and androsten-3-beta-17-beta-diol-disulfate 1 were influenced by *LTK* (p=6e-09), *DSE* (p=1e-07), and *PLAC4* (p=4e-09,) respectively.

For the diet-related metabolites heptanoate, adipate, azelate, sebacate, dodecanedioate, and glutarate, the gene *CAPN9* (p=1e-7) influenced heptanoate in the AA population. In the EA causal network, the gene, *MYO1A*, influenced (p=4e-09) heptanoate and the gene *PDE4DIP* influenced dodecanedioate (p=1e-14). Among 16 long chain fatty acids in the analysis, oleate and nonadecanoate, were affected by LoF loci in the AA causal network. Oleate was affected by two genes, *GPR97* (p=4e-07) and *MAP10* (p=4e-9), and the nonadecanoate was affected by *CLEC4C* (p=4e-10). In the EA causal network, long chain fatty acid myristoleate was affected by two genes, *ZNF211* (p=1e-9) and *EGFL8* (p=8e-13).

In the module including amino acid related metabolites in the AA causal network, two peptides from the gamma-glutamyl sub-pathway were influenced by LoF loci: leucine by *OR11G2* (p=5e-10) and *CYP2A6* (p=1e-11), and glutamate by *COASY* (p=4e-09) and *STPG1* (p=1e-12). In the EA population, no LoF mutation had any impact on metabolites from the amino acid module.

**Findings in EA and AA populations based on metabolite pathways:** In AA, 34 metabolites were significantly affected by LoF loci; in EA, 19 metabolites. A summary of the pathway-based metabolites is provided in Table 3.

**Table** 3. Number of pathway-based metabolites influenced by LoF loci.

| Pathway | AA | EA |
|---|---|---|
| Lipid | 11 | 13 |
| Amino Acid | 15 | 3 |
| Peptide | 5 | 1 |
| others | 3 | 3 |

Figure 6 shows the minor allele frequency (MAF%) for LoF loci and the selected LoF loci in each of the populations. This figure does not show any differences in MAF% between AA and EA populations.

To validate the findings, we used a validation set from the same AA population including 672 samples with 154 LoF carriers with MAF>7. In the replication analysis, 8 of the gene-metabolite findings were replicated (p <2.6E-06), depicted with an asterisk in Table 2.

## Discussion

To make fully informed inference about the likely function of genomic variants, systems biology approaches that help to understand the pathway of the genetic effects on disease through intermediate molecular levels offer improvements over standard single-variable analyses commonly used in GWAS. We investigated the effects of LoF genomic variants on the human metabolome in the two ethnic groups of EA and AA participants from the ARIC study. To investigate LoF-metabolite relationships, we selected genetic variants that were found to be statistically associated with metabolites by both a penalized model (CCRS) and single variant tests. We identified metabolomic causal networks using the G-DAG algorithm that is based on a systems biology method. This allowed us to integrate genomic and metabolomic relationships, built on directed acyclic graphs that portray directions, thus meeting the assumptions of the structural equation modeling that we used to assess the LoF-metabolite findings. Because differences between EA and AA participants, such as the size of covariate effects, different geographical regions and as a result different diet, could impact biochemical pathways, we avoided any comparisons of EA and AA metabolomic causal networks in this study.

Although many of our findings were for rare genetic variants, which made it difficult to corroborate our results from published reports, we were able to find published reports that supported some of our results. We identified some LoF mutations with high impact on fatty acid metabolism that may play a role in the pathogenesis of metabolic syndrome and cardiovascular disease. For instance, KIAA1755 has been found to be associated with heart rate from exome chip meta analyses (29). We found a strong relationship between KIAA1755 and eicosapentaenoate ($p<5e-14$). The metabolite eicosapentaenoate is related to essential hypertension, which is the most common type of hypertension with no known cause (30)(31). From the identified metabolomic causal network based on Mendelian randomization principles, we found the metabolite eicosapentaenoate was among 4 metabolites with high impact on arachidonic acid (3). It has also been validated clinically that arachidonic acid has the greatest positive impact on triglyceride levels, a risk factor of cardiovascular disease (26). To assess the effect of triglycerides on essential hypertensive patients, in a separate previous clinical study, 900 patients were examined and a link was found between increased plasma triglyceride levels with more fatal events in essential hypertensive patients (32). These relationships are depicted in Figure 7 and may provide some insights into the disease process. The gene *KIAA1755* is a gene/protein of unknown function and findings here may reveal new avenues into the gene function and the understanding of the disease etiology.

We found a significant influence of PDE4DIP on the metabolite dodecanedioate from a lipid pathway. Dodecanedioate supplementation in non-insulin-dependent diabetic patients (but not healthy controls) reduced glucose levels without altering insulin levels (33). Another interesting finding was that the gene *MYO1A* was found to have a strong impact on the metabolite heptanoate from a lipid pathway. This metabolite has been used as a treatment for Long-chain L-3 hydroxyacyl-CoA dehydrogenase deficiency, a condition in which the body is unable to break down certain fats (34).

In Table 2, we showed significant associations between *CLDN17* with increased levels of isoleucine (pvalue=2e-12), and *APOA1BP* (pvalue=9e-13) with increased levels of leucine, in the AA population. The relationships were through heterozygous LoF variants, suggesting that haploinsufficiency of these genes results in an increase of the corresponding metabolites. Figure 8 illustrates that these two metabolites were highly connected in the causal network, meaning that their effects can spread across the metabolomic system by influencing multiple metabolites directly and indirectly. Note that all participants in AA population were from Jackson, Mississippi, in an attempt to control for environmental confounders, such as regional dietary variations that can influence the metabolome.

Leucine and isoleucine are branched-chain essential amino acids. Studies have shown that higher levels of leucine or isoleucine are strongly related to insulin resistance, obesity, and higher risk of type 2 diabetes (35, 36). We found a strong association between the gene *CLDN17* (Claudin-17) and isoleucine. This gene is a tight junction protein that facilitates anion-selective paracellular transportation and is predominantly expressed in proximal nephrons (37). There is evidence suggesting the electrogenic transport of amino acids including leucine and isoleucine by proximal tubules (38). *APOA1BP* is an apolipoprotein A-I (ApoA-I) binding protein, which is essential in maintaining cholesterol homeostasis and plays a role in cardiovascular disease (39). Low levels of ApoA-I and HDL cholesterol are linked to risk of cardiovascular events (40), whereas increases in several amino acids, including leucine, are linked to low levels of HDL cholesterol and insulin resistance in patients with renal dysfunction(41).

In the AA population, we found the metabolite gammaglutamylleucine from the peptide pathway was influenced by a LoF mutation in *CYP2A6* (p-value=1e-11). The other related findings were relationships between *CYP2A13* and the metabolites pyroglutamine (p-value = 5.13e-07) and hydroxyphenyllactate (pvalue=1.81e-08). The LoF mutation in *CYP2A13* revealed a pleiotropic action on two metabolites, pyroglutamine and hydroxyphenyllactate. *CYP2A6* and *CYP2A13* are both members of cytochrome P450 superfamily of enzymes, which are involved in many catalytic reactions and drug metabolism including nicotine metabolism. Genetic variants in these genes present a lower risk of developing tobacco-related lung cancer in multiethnic populations including African-American smokers (42–44). An increased level of gammaglutamylleucine has been identified as an indicator of anti-obesogenic metabolism and the strongest risk-decreasing predictor of death (41). Additionally, the inhibition of *CYP2A6*- and *CYP2A13*-mediated metabolism of nicotine can increase the activity of gammaglutamylcysteine syntheses that results in the overproduction of pyroglutamate (40). It has been proposed that pyroglutamate has a xenobiotic detoxifying property (45). The second metabolite associated with *CYP2A13* was hydroxyphenyllactate, which is a tyrosine metabolite, and the L-form of it has been reported to be highly elevated in the urine of patients with pheylketonuria and tyrosinemia. Taken together, using our approach, we identified two separate metabolites gammaglutamylleucine and pyroglutamine that had significant associations with LoF variants of two genes, *CYP2A6* and *CYP2A13,* respectively from the same family of genes. Thus, these metabolites could be functionally related given their protective functions. Furthermore, our analysis method was able to uncover the potentially pleiotropic effects of the gene *CYP2A13* on two different metabolites, pyroglutamine and hydroxyphenyllactate, that may act independently of each other. Nonetheless, further study is required to understand the biological functions of these genes on related metabolites and diseases. Identifying pleiotropic genes reveals important biological relationships among molecular components or clinical phenotypes and leads to understanding of complex biological mechanisms, disease pathogenesis and underlying co-morbidities (46)(47). This understanding may improve and speed drug development and furthermore, predict possible side effects (48)(49).

Although systems approaches to molecular pathway identification will not replace mechanistic experiments, they are complementary and hypothesis-generating, especially in the era of large scale data to narrow the search space, identify targets for intervention, and define pathways that spread the effect of any intervention in the system. The integrated approach introduced here facilitates a mechanistic understanding through incorporating and modeling relationships of a large number of metabolites in genome analysis, improves genome pathway identification, and identifies genes with pleiotropic actions. The focus of the current study was on LoF variants. For future studies, we aim to extend the work to additional previously-identified associated genetic variants (7) and utilizing new clustering approaches (50)(51).

## Conclusions

Metabolomic data function as an intermediate at the molecular level to illuminate mechanisms underlying a specific genetic variant, to identify biological pathways linking the genome to disease, and to discover valuable clinical biomarkers. By integrating results of the genome analysis with metabolomic relationships, and as a result, improving the focus on biological systems, including the identification of pleiotropic gene actions, we improved genome analysis and studied how naturally occurring genetic variants can affect metabolites.

### Methods

**Metabolomic and genomic data:** The data was from the ARIC Study, which enrolled 15,792 middle-aged individuals (45 to 64 years) at baseline. For more details of the cohort characteristics see https://www2.cscc.unc.edu/aric/system/files/ CohortCharacteristics.pdf and Supplementary, table S1. Serum metabolomic and genomic data were available on 1376 individuals from the AA and EA populations in the study. The dbGAP accession number for ARIC study data is phs000668.v3.p1. A total of 602 metabolites were detected and semi-quantified by Metabolon Inc. (Durham, North Carolina). After excluding metabolites with at least 50% missing values across all samples, metabolites with unknown chemical structures, and metabolites (or any transformation of them) that did not follow a normal distribution, we focused on 122 named metabolites, Supplementary table S8. Details of metabolite assessments are provided in the next section and Supplementary, section Metabolite measurement. Replication samples were available; and we utilized them to identify and impute the missing values with an optimal approach (52). Since covariates could have a profound effect on metabolites, we adjusted metabolites for age, gender, and body mass index, phase (two different time points that the metabolites were measured), and ten principal components (to adjust for population stratification) using a linear regression. In addition, we carried out the analysis for AA ad EA populations separately. LoF variants were defined as sequence changes caused by single nucleotide variants or small insertions and deletions. The exomes were sequenced on the individuals and 372 and 451 LoF variants were identified for AA and EA populations respectively (23). Variants were annotated using ANNOVAR (53) and dbNSFP v2.0 (23) according to the reference genome GRCh37 and National Center for Biotechnology Information Reference Sequence.

After exome capture with VCRome 2.1 (NimbleGen, Inc., Madison, WI), sequencing was carried out using Illumina HiSeq instruments. Using Burrows-Wheeler aligner (54), sequences were aligned to Genome Reference Consortium Human Build 37. Allele calling and variant-call file construction were carried out with the Atlas2 suite (Atlas-SNP and Atlas-Indel). Single-nucleotide variants (SNVs) were removed using the following criteria: SNVs with a SNP posterior probability less than 0.95, a total depth of coverage less than 6×, an allelic fraction of <0.1, fewer than three variant reads, 99% reads in a single direction and homozygous reference alleles with <6× coverage. More stringent filtering was applied on low-quality single-nucleotide substitutions with total depth less than 10; low-quality indels with the following differences: allelic fraction <0.2 for heterozygous variants and <0.8 for homozygous variants, minimum total depth less than 60, and variant reads smaller than 30.
All LoF varients in this study were confirmed to have a premature stop codon in a coding exon, disruption of an essential splice site, or an indel predicted to disrupt the downstream reading frame (23).

Serum metabolomic and genomic data were available on an additional subset of individuals from the same population, 672 African-Americans, which were used to validate the findings in AA population.

**Metabolomic measurement**
Metabolic profiling was carried out on fasting serum samples from the baseline examination stored at -80°C. Metabolites were measured using untargeted gas chromatography-mass spectrometry and liquid chromatography-mass spectrometry (GC-MS and LC-MS-based quantification protocols by Metabolon, Inc., Durham, NC). Because the laboratory work for this study is expansive and spanned many days, a data normalization step was performed to correct for variation resulting from differences in instrument tuning from day to day. Metabolites were identified by comparison to library entries of purified standards or recurrent unknown entities. Several types of internal controls were analyzed in concert with the experimental samples. Tables S2 and S3 in supplementary describe these quality assurance and quality control samples and standards. Further QC consisted of four major components: the LIMS, the data extraction and peak-identification software, data processing tools for QC and compound identification. The proposed methods, although robust, assume normality of the metabolomics and risk factor data. Log transformation may not be the best transformation for all metabolites. Different transformations were assessed, such as square, root square, log etc., and different metabolites were transformed to a normal distribution using different transformations.

**Genome analysis**: To select the genetic variants associated with metabolites, we applied

1- Convex-Concave Rare variant Selection (CCRS). The CCRS approach (24) is a penalized model with some constraints on the design matrix and coefficient vector to provide parsimony of selected covariates, similar to lasso penalization. Using this approach, we had a multivariable regression model which included all LOF variants in the model at a time. Therefore, in contrast to single variant tests, the CCRS approach is a multiple regression method that simultaneously selects the most promising genome variants and estimates their effects on metabolites of interest. Applying the CCRS not only increases the power of identification due to avoiding multiple comparison adjustments, it also takes into account local linkage disequilibrium and prevents overestimation.

2- In addition to the CCRS, to identify mutations in the coding sequence with significant effect on metabolites, we sought association with the metabolites using a linear regression model for each variant at a time. For rare variants, we aggregated their effects by summing the number of LoF variants in each gene and then tested the association of this sum with metabolites by linear regression.

3- We then focused on the variants that were commonly selected by the two steps above to reduce the false discovery rate.

**Metabolomic causal networks**: Metabolomic causal networks represent metabolites as nodes connected by directed edges indicating the relationships among metabolites. A missing link between two metabolites in the causal network means no relationship, and a link between two metabolites represents the relationship after conditioning on the effect of other metabolites in the analysis (conditional analysis). In a causal network, directions represent cause and effect relationships in observational studies identified based on Mendelian principles/instrumental variables, using variation in the system that is free of confounding (55)(56)(57). The G-DAG (genome directed acyclic graph) algorithm (21)(25) with clinically validated novel findings (26) is based on the principle that the genome inherited variation is a causal factor of metabolomic changes and not the other way around. Some of applications of the G-DAG algorithm are (3)(20)(21)(25)(58). The G-DAG algorithm was employed to identify metabolomic causal networks, see below for the steps of the G-DAG algorithm. To enforce the assumptions of instrumental variable application and identify robust directions among metabolites, the G-DAG algorithm has the following features (58):

1. Extracting information from multiple genetic variants using principal component analysis to create strong instrumental variables;
2. Independent instrumental variables due to application of principal component analysis;
3. Multiple independent instrumental variables used for each metabolite, to make overall instrumental variables even stronger.

For identification of the AA and EA metabolomic causal networks, in total 353 and 412 instrumental variables were utilized, respectively. These instrumental variables that were generated from whole exome sequencing data represent the global genomic effects on metabolites despite an individual genetic variant effect. Figure S1 in Supplementary represents the outcomes of the G-DAG algorithm for EA and AA populations.

To obtain the best fit for the networks, we employed structural Hamming distance (59), a well-established assessment for the quality of fit in networks, e.g. see (60)(61). The distance is a function of sample size, variables, and neighbors; a smaller distance indicates a better fit. We considered a set of tuning parameters including 0.0005, 0,001, 0.005, 0.01, 0.05. For the setting with alpha 0.001 and 0.005 the average of the distance on 45 replications was minimum. Therefore, we carried out the analysis at statistical significance level 0.001.

**The G-DAG Algorithm**

The G-DAG algorithm (25) uses principal components (generated from genome variation) as strong instrumental variables to identify a stable network over the metabolites. Note that the primary aim was identification of a causal network over the set of variables of interest (here metabolites) and the genome information is used as a tool to aid in identifying directionality among the metabolites.

**First**, we start by reducing the number of SNPs by considering the fact that some SNPs are nearly perfectly correlated (>0.80) with others, so that one SNP can thereby serve as a proxy for many others in the analysis. To determine a proxy, we use hierarchical clustering and a measure of linkage disequilibrium (62).

**Second**, the genomic information was summarized using principal component analysis.

**Third**, the set of principal components responsible for more than 90% of the variation was selected.

**Fourth**, the tuning parameter in the G-DAG algorithm was set using the structural Hamming distance, which is a function of variables, sample size and neighboring.

**Fifth**, in a constraint-based algorithm, the G-DAG identified a causal network over the principal components as instrumental variables and metabolites.

**Pleiotropy identification**: The pleiotropic effect of an inherited gene can refer to a single nucleotide polymorphism (SNP), an entire gene, a large segment of the genome containing multiple genes (63), or regulatory motifs across the genome (64). We point out the definition of pleiotropy considered in this study as genes with more than one functionality, the ability of a gene to cause distinct phenotypic traits (65). The word "distinct" should not be interpreted as independent since some dependency can arise from shared environmental influences and direct physiologic and biochemical relationships (66). Alternatively, in association studies, where one trait at a time is analyzed, a gene associated with more than one trait cannot be identified with pleiotropic action since this relationship might be due to association between the two traits and not different functionality of the gene. We assessed the pleiotropic action through identification of the genes with direct effect on more than one metabolite.

When a study has a large number of traits, such as metabolites, illumination of underlying relationships provides an opportunity to improve the power of genome analysis, and consequently identification of pleiotropic gene actions. Using causal networks established in principles of Mendelian randomization and application of structural equation modeling, we identified pleiotropic actions by removing findings that are due to metabolomic relationships and not the gene's direct effect on metabolites.

**Structural equation modeling**: Here, structural equations were used to model genetic variants with pleiotropic effects based on the relationships of genetic variants and metabolites. In the general form, with *M* representing metabolites and *G* representing genetic variants, the structural equations are

$$M_i|AM(K_R) = \sum_{j=h+1}^{l} \lambda_{ij} M_j + U_i$$

$$U_i|AM(K_R) = \sum_{m=1}^{h} \gamma_{im} G_m + e$$

where $m = 1,\ldots, h$ ; $j = h+1,\ldots, l$ ; $i = l+1$; and $\lambda_{ij} \neq 0$ is equivalent with (j ⟶ i) and $\gamma_{im} \neq 0$ is equivalent with ($m \longrightarrow i$). The notation $AM(K_R)$ is called casual parameter (55). It stands for Assignment Mechanism given the Knowledge about Response (67). To assure that the assumptions of structural equation modeling were met, we first identified the metabolomic causal networks. The causal networks are illustration of Assignment Mechanisms behind observations (67) (68) which can be illuminated by gathering any Knowledge about Response variables. The reason for conditioning the equations above on the causal parameter is to show that the equations are written based on an identified causal network that meet the required assumption. Note that in a metabolomic causal network, each metabolite is a response variable and the causal network represents metabolites that were involved in assigning value to each response variable (a metabolite). For applications of structural equation modeling see (27)(69).

**Declarations**

Abbreviations

AA: African-American/non-white

EA: European-American/white

ARIC study: Atherosclerosis Risk in Communities

G-DAG: Genome Directed Acyclic Graph

CCRS: Convex-Concave Rare Variant Selection

LoF: Loss of Function

SEM: Structural Equation Modeling

Ethics approval and consent to participate: Not applicable
Consent for publication: Not applicable
Availability of data and material: The dbGAP accession number for ARIC study data is phs000668.v3.p1.
Competing interests: The authors declare that they have no competing interests.
Funding: The first author, Dr. Yazdani was supported by a training fellowship from the Keck Center for Interdisciplinary Bioscience Training of the Gulf Coast Consortia (Grant No. RP140113). DJ Schaid was supported by the U.S. Public Health

Service, National Institutes of Health, contract (Grant No. GM065450). MR Kosorok was supported by National Cancer Institute grant P01 CA142538. The funding was for the design and writing of the manuscript.

Authors' contributions:

AY1 introduced, implemented, and applied the methods, interpreted the results and wrote the manuscript.
AY2 collaborated in data analysis, writing the manuscript and interpreting the results.
SHE, DJS, and MRK provided essential comments and suggestions to improve the manuscript.
GD collaborated on interpretation of the findings.
AS collaborated on algorithm implementation and visualization.
All authors have read and approved the manuscript.

Acknowledgements: The corresponding author, Dr. Yazdani, appreciates Dr. Boerwinkle for providing the data and introducing the pleiotropic topic to her. Dr. Yazdani also thanks the staff and participants of the ARIC Study for their important contributions. ARIC study is supported by the U.S. Public Health Service, National Institutes of Health, contract grant number GM065450.

## Figure Legends

**Figure 1.** The steps to conduct gene-metabolite investigation. Abbreviation: SEM: structural equation modeling; CCRS: convex-concave rare variant selection; G-DAG: genome directed acyclic graph. Details of the CCRS method (for the genome analysis) and the steps of the G-DAG algorithm (for constructing metabolomics causal networks) are described in the methods section.

**Figure 2**. *GPR97*-metabolite pathway. The red cross represents the pathway that is not significant after modeling the correlated metabolites using structural equation modeling. In the figure, glycocholenate stands for glycocholenate-sulfate.

**Figure 3**. *BNIPL*-metabolite pathway. The red cross represents that the pathway was not significant after modeling the relationship between the metabolites octanoylcarnitine and decanoylcarnitine.

**Figure 4**. The module including metabolites related to gut metabolism. Urea and glutarylcarnitine are neighbors of the module. Metabolites influenced by LoF mutations in AA population are depicted in larger scale.

**Figure 5**. Hormone related metabolites as a module and pyroglutamine as a neighbor of the module. The metabolites influenced by LoF mutations in AA population are depicted in larger scale. Abbreviation: aS1/2 stands for andersten-3-beta-17-betadiol-disulfate1/2. The capital S at the end of glycocholenateS, pregnendioldiS, hydroxypregnenolonediS, and pregnsteroidmonoS stands for sulfate.

**Figure 6**. Left panel: MAF<20% for LoF mutations. Right Panel: MAF% for LoF mutations identified with significant impact on metabolites. The x-axes indicate the alternative allele, and the y-axes indicate the MAF%.

**Figure 7**: A pathway from the genome to metabolic disorder and cardiovascular disease through metabolomics and the risk factors. Yellow represents metabolites. Red indicates risk factors of cardiovascular disease and components of metabolic disorder. Abbreviations: TG: Triglycerides, EPA: Eicosapentaenoate, EH: Essential Hypertension, Docosapentaenoyl-G: Docosapentaenoyl-Glycerophosphocholine

**Figure 8**. The impact of leucine and isoleucine on metabolomics. Leucine and isoleucine influence multiple neighboring metabolites, broadly influencing the system.

Supplementary files

1. Figure S1. European-American metabolomic causal network using the G-DAG algorithm.
2. Table S1. Characteristics of ARIC Cohort
3. Metabolite assessment and Tables S2 and S3 for QC.
4. Tables S4 and S5. LoF-metabolite relationships using the CCRS approach for EA and AA populations respectively.
5. Tables S6 and S7. LoF-metabolite relationships using the single variant test for EA and AA populations respectively.
6. Table S8. List of 122 named metabolites measured in ARIC study

**Table 1**. EA (European-American) population, LoF-metabolite relationships at p-value 4e-07 or smaller. Effect sizes measured in standard deviation units to facilitate comparison. The last column "Human phenotypes related to metabolites" was obtained through Human Metabolome database, http://www.hmdb.ca/.

| Gene | SNP | MAF | Metabolite | Path-way | Effect Size (Std. | p-value | Human phenotypes related to metabolites |
|---|---|---|---|---|---|---|---|
| ***TPTE*** | X21:10907041:C:A | 8 | O-sulfo-L-tyrosine | - | 1.75 (0.34) | 2e-07 | |
| | | | myoinositol | Lipid | 2.23 (0.35) | 2e-10 | Schizophrenia |
| ***MYO1A*** | X12:57441459:G:A | 9 | heptanoate | Lipid | 1.95 (0.33) | 4e-09 | Cardiac disease |
| | | | glycerol | Lipid | 1.70 (0.33) | 3e-07 | Schizophrenia, Diabetes mellitus type 2 |
| ***GZMM*** | X19:549171:C:T | 4 | octanoylcarnitine | Lipid | 1.46 (0.26) | 4e-08 | Medium-chain acyl-CoA dehydrogenase deficiency |
| ***PLAC4*** | X21:42551432:T:A | 9 | androsten-3-beta-17 beta-diol-disulfate1 | Lipid | 1.95 (0.33) | 4e-09 | NA |
| ***LRTOMT*** | X11:71807767:A:C | 10 | betaine | | 0.67 (0.23) | 7e-09 | Chronic renal failure and hemodialysis, Schizophrenia |
| ***PDE4DIP*** | X1:145074975:G:A | 17 | dodecanedioate | Lipid | 1.84 (0.24) | 1e-14 | Diabetes mellitus type 2 (34) |
| ***OBSCN*** | X1.228469903.A.T | 8 | serine | Amino acid | 1.89 (0.34) | 7e-08 | Schizophrenia and epilepsy, Leukemia, Heart failure |
| ***THSD7B*** | X2:138030234:C:T | 19 | glycylvaline | Peptide | 1.61 (0.23) | 1e-12 | NA |
| ***C5orf45*** | X5:179280392:G:A | 8 | cis-4-decenoylcarnitine | Lipid | 3.10 (0.34) | 2e-16 | Celiac disease, Very long chain acyl-Coa dehydrogenase deficiency |
| ***DSE*** | X6.116600895.G.A | 10 | pregn steroid monosulfate | Lipid | 1.65 (0.31) | 1e-07 | Schizophrenia |
| ***EGFL8*** | X6:32134395:C:G | 8 | myristoleate | Lipid | 2.51 (0.35) | 8e-13 | NA |
| ***GPNMB*** | X7:23313823:G:T | 12 | lactate | Amino acid | 1.80 (0.28) | 3e-10 | Hepatobiliary, Psychiatric, Mitochondrial dysfunction, Metabolism and nutrition disorders |
| ***C10orf53*** | X10:50901917:C:A | 8 | hydroxybutyrate | Amino acid | 1.78 (0.35) | 4e-07 | Schizophrenia, Pyruvate dehydrogenase deficiency |
| ***SERPINA10*** | X14:94754643:C:T | 14 | prohydroxyproline | | 1.45 (0.26) | 4e-08 | NA |
| ***LTK*** | X15:41796352:C:A | 8 | glycocholenatesulfate | Lipid | 2.04 (0.35) | 6e-09 | NA |
| ***CCDC154*** | X16:1484536:C:T | 14 | glycerol-phosphorylcholine | Lipid | 1.41 (0.26) | 1e-07 | Multi-infarct dementia |
| ***RHBDL1*** | X16:726189:C:T | 14 | phosphate | Lipid | 1.75 (0.26) | 4e-11 | NA |
| ***ZNF211*** | X19:58153465:T:A | 12 | myristoleate | Lipid | 1.73 (0.28) | 1e-09 | NA |

**Table 2**. AA (African-American/non-white) population LoF-metabolite relationships at p-value 1e-6 or smaller. Effect sizes measured in standard deviation units to facilitate comparison. The column, "Human phenotypes related to metabolites" was obtained through Human Metabolome database, http://www.hmdb.ca/. LCAC: long-chain acylcarnitine. The last column represents the p-value in the replication analysis. Genes with asterisk are replicated.

| Gene | SNP | MAF | Metabolite | Path-way | Effect Size (Std.Erro | p-value | Metabolite related human phenotypes | Rep-lication p-value |
|---|---|---|---|---|---|---|---|---|
| ***GPR97*** | X16:57707232:G:C | 4 | glycocholenate-sulfate | Lipid | 1.09 (2.34) | 3e-11 | _ | NA |
| | | | oleate | Lipid | 1.79 (0.35) | 4e-07 | LCAC accumulation, Inflammation, Schizophrenia, Gestational diabetes | |
| ***GGN*** | X19:38875072:G:A | 8 | androsten-3-beta-17-beta-diol-disulfate 1 | Lipid | 1.68 (0.27) | 1e-09 | - | NA |
| | | | xanthine | Nucleotide | 1.63 (0.28) | 4e-09 | Nervous system disorders, Renal failure | |
| ***CYP2A13*** | X19:41594954:C:T | 6 | pyroglutamine | Amino acid | 1.67 (0.33) | 5e-07 | Nervous system, and Metabolism & nutrition disorders | NA |
| | | | hydroxyphenyllactate | Amino acid | 1.87 (0.33) | 2e-08 | Supradiaphragmatic malignancy. Liver dysfunction | |
| ***CLDN17*** | X21:31538461:G:A | 8 | isoleucine | Amino acid | 1.96 (0.29) | 8e-12 | Heart failure, Leukemia, Maple syrup urine disease | NA |
| | | | glycerol | Lipid | 1.72 (0.29) | 2e-09 | Schizophrenia, Diabetes mellitus type 2 | |
| ***CYP2A6*** | X19:41351363:T:A | 6 | gamma-glutamylleucine | Peptide | 1.92 (0.28) | 1e-11 | NA | NA |
| | | | Gamma-glutamylthreonine | Peptide | 1.27 (0.30) | 1e-10 | - | |
| ***ZSWIM1*** | X20:44511257:G:A | 7 | Phenylacetyl-glutamine | Amino acid | 2.75 (0.35) | 4e-15 | NA | NA |
| ***CRYBB3*** | X22:25599863:G:T | 3 | trans-4-hydroxyproline | Amino acid | 2.19 (0.37) | 6e-09 | Renal disorder | NA |
| ***STPG1*** | X1:24727815:G:T | 3 | gamma-glutamyl-glutamate | Peptide | 2.49 (0.35) | 1e-12 | NA | NA |
| ***CLSPN*** | X1:36208741:C:T | 7 | urea | Amino acid | 2.13 (0.28) | 1e-13 | Infection | NA |
| ***CLEC4C*** | X12:7899913:C:A | 4 | nonadecanoate | Lipid | 2.36 (0.37) | 4e-10 | NA | NA |
| ***BNIPL*** | X1:151016171:G:A | 3 | decanoylcarnitine | | 2.17 (0.35) | 7e-10 | Medium-chain acyl-CoA dehydrogenase deficiency | NA |
| ***MAP10*** | X1:232942469:G:A | 6 | oleate | Lipid | 2.07 (0.35) | 4e-09 | LCAC accumulation, Inflammation, Schizophrenia, Gestational diabetes | NA |
| ***OR11G2*** | X14:20666175:C:A | 22 | gamma-glutamylleuci | Peptide | 1.11 (0.18) | 9e-10 | NA | 0.03 |
| ***FAM151A*** | X1:55075006:G:A | 6 | pregnendiol-disulfate | Lipid | 2.59 (0.35) | 2e-13 | Schizophrenia | NA |
| ***APOA1BP*** | X1:156563265:C:T | 4 | leucine | Amino acid | 2.51 (0.35) | 9e-13 | Heart failure, Leukemia, Maple syrup urine disease | NA |
| ***KIAA1755*** | X20:36869005:G:A | 8 | eicosapentaenoate | Lipid | 2.08 (0.27) | 5e-14 | Essential hypertension | NA |
| ***ELSPBP1*** | X19:48523114:G:A | 4 | erythritol | Xenobiotics | 2.52 (0.38) | 2e-11 | Pentose phosphate pathway abnormalities | NA |
| ***COASY*** | X17:40717487:A:G | 8 | gamma-glutamylgluta | Peptide | 1.76 (0.29) | 4e-09 | NA | NA |
| ***CD300C*** | X17:72540958:G:A | 4 | docosahexaenoylglycero- | Lysolipid | 2.11 (0.30) | 2e-12 | | NA |
| ***C16orf55*** | X16:89724661:G:T | 11 | androsten-3-beta-17-beta-diol-disulfate 2 | Lipid | 1.98 (0.24) | 2e-16 | NA | NA |
| C15orf32 | X15:93015466:A:T | 17 | glycerol-3-phosphate | Lipid | 1.11 (0.19) | 4e-09 | NA | 0.9 |
| LTK | X15:41799325:G:A | 18 | urea | Amino acid | 1.14 (0.19) | 6e-09 | Infection | 0.6 |
| AVEN | X15:34159987:G:A | 4 | urea | Amino acid | 1.71 (0.30) | 2e-08 | Infection | NA |
| UPK2 | X11:118827917:G:A | 7 | glycerol | Lipid | 1.89 (0.30) | 3e-10 | Schizophrenia, Diabetes mellitus type2 | NA |
| SLC25A27 | X6:46623768:G:A | 21 | O-sulfo L-tyrosine | - | 1.24 (0.20) | 1e-09 | NA | NA |
| DCLK3 | X3:36756821:A:C | 10 | phenylacetylglutamine | Amino acid | 1.62 (0.28) | 4e-09 | NA | NA |
| TMPRSS3 | X21:43792873:A:G | 11 | phenylacetylglutamine | Amino acid | 2.04 (0.25) | 2e-16 | NA | NA |
| GUCA1C | X3:108672558:C:A | 6 | glutarylcarnitine | Amino acid | 2.32 (0.35) | 4e-11 | NA | NA |
| NWD1* | X19:16908698:G:A | 22 | carboxy4methyl5propyl2furanpropanoate | Lipid | 0.97 (0.10) | 1e-06 | Psychiatric, kidney, cardiovascular and gastrointestinal disease, seizure | 1e-8 |
| POLR1E* | X9:37495945:C:A | 21 | glycerol3phosphate | Lipid | 1.50 (0.23) | 2e-10 | Multi-infarct dementia | 3e-20 |
| P2RX7* | X12:121570899:G:T | 8 | tyrosine | Amino acid | 1.59 (0.32) | 7e-07 | Leukemia, Hypothyroidism, Myocardial infection, Schizophrenia, | 1e-7 |
| CAPN11* | X6:44147355:G:A | 14 | aminobutyrate | Amino acid | 1.18 (0.23) | 7e-07 | Metabolism, hepatic disorder, Epilepsy, Aciduria, Febrile seizures, | 1e-7 |
| OR11H4* | X14:20711121:T:A | 22 | catecholsulfate | Xenobiotics | 1.24 (0.19) | 2e-10 | NA | 1e-7 |
| C6orf25* | X6:31692558:C:T | 10 | leucine | Amino acid | 1.85 (0.31) | 7e-09 | Heart failure, Leukemia, Maple syrup urine disease, Schizophrenia, Epilepsy | 8e-13 |
| | | | isoleucine | | 1.79 (0.27) | 1e-10 | | 5e-14 |
| | | | valine | | 1.72 (0.20) | 3e-09 | | 2e-17 |